\documentclass[onecolumn,prb,groupedaddress,preprintnumbers,amsmath,amssymb, floatfix]{revtex4}
\usepackage{amsmath}
\usepackage{graphicx}
\usepackage{dcolumn}

\begin{document}
\title{Phase behavior of ionic liquid crystals}

\author{S. Kondrat, M. Bier, and L. Harnau$^*$}
\affiliation{
         Max-Planck-Institut f\"ur Metallforschung,  
         Heisenbergstr.\ 3, D-70569 Stuttgart, Germany, 
         \\
         and Institut f\"ur Theoretische und Angewandte Physik, 
         Universit\"at Stuttgart, 
         Pfaffenwaldring 57, 
         D-70569 Stuttgart, Germany
	 }
\email[]{E-mail: harnau@fluids.mpi-stuttgart.mpg.de} 
\date{\today}

\begin{abstract}
Bulk properties of ionic liquid crystals are investigated using density 
functional theory. The liquid crystal molecules are represented by 
ellipsoidal particles with charges located in their center or at their 
tails. Attractive interactions are taken into account in terms of the 
Gay-Berne pair potential. Rich phase diagrams involving vapor, isotropic 
and nematic liquid, as well as smectic phases are found. The dependence 
of the phase behavior on various parameters such as the length of the 
particles and the location of charges on the particles is studied. 
\end{abstract}
\maketitle

\section{Introduction}
The transportation of charges \cite{adam:94,neil:03} and ions \cite{kato:02,yosh:02}
in liquid crystals has attracted much attention because these materials are 
expected to serve as anisotropic conductors due to their self-organized structures. 
For this purpose the design and control of molecular interactions and 
microphase-segregated structures in ionic liquid crystals is essential. Moreover, the 
macroscopic orientation of self-organized monodomains plays an important role in the 
improvement of the conducting properties because the boundaries in randomly oriented 
polydomains disturb high and anisotropic transportation of charges and ions.
Ionic liquid crystals can also be used as ordered solvents or organized 
reaction media. In these anisotropic solvents other chemo- and regioselectivities 
besides those in conventional solvents can be obtained for several types of reactions.

Various types of ionic liquid crystals have been prepared. The ionic liquid crystals 
based on imidazolium or pyridinium salts containing weakly coordinating anions
such as BF$^-_4$ and PF$^-_6$ are representative due to their thermal and 
electrochemical stabilities. \cite{gord:98,brad:02,lee:03,roch:03,ster:07,kouw:07} 
In these and related materials the liquid crystalline phases are induced by microphase
segregation of ionic moieties of long alkyl or perfluoroalkyl chains. The types of 
liquid crystalline phases depend on the molecular shape and location of the 
ionic parts on the molecules. Many ionic molecules containing single side
chains form smectic structures. The influence of the anion type and chain length 
on the liquid crystalline phases has been investigated for 1-alkyl-3-methylimidazolium
salts. \cite{holb:99,hard:01,brad:02} 
Very recently an efficient synthetic route towards calamitic guanidinium salts 
has been developed. \cite{saue:09}  These guanidinium salts exhibit stable 
mesophases. Metal-based ionic liquid crystals containing a tetrahalometalate ion and 
$N$-alkylpyridinium salts exhibit a variety of liquid crystalline phases 
ranging from smectic to columnar or even cubic phases. \cite{neve:98,neve:01}
These metal-based materials can exhibit interesting properties as metal complexes 
such as chromism, magnetism, polarizability, redox behavior, and catalysis.

Thermotropic columnar liquid crystalline phases are formed by self-organization 
of fan-shaped imidazolium molecules. \cite{yosh:04a} In these phases the 
imidazolium parts form one-dimensional paths inside the columns. These columnar 
materials are macroscopically aligned by shearing on the glass substrate. 
Taubert described the use of an ionic liquid crystal as a template to synthesize 
CuCl nanoplatelets. \cite{taub:04,taub:05} The platelets formed in the mesophase 
are relatively large and interconnected, whereas smaller platelets without permanent
junction were formed in the isotropic liquid. Moreover, it has been shown that 
ionic liquids containing a guanidinium moiety tethered to a pentaalkyloxytriphenylene 
unit form platelike structures and columnar mesophases. \cite{saue:08}

In a recent review, Binnemanns has discussed many experimental studies that have
been devoted to the synthesis and properties of ionic liquid crystals. \cite{binn:05} 
In particular he comes to the conclusion that theories that can explain the 
influence of the anisotropic charge distribution on the mesophase stability on 
ionic liquid crystals are still lacking. The aim of the present paper is to provide 
theoretical insight into the underlying mechanisms responsible for the formation of 
bulk liquid crystalline phases in ionic liquids. In Sec. II we define the system 
under consideration, and we describe the density functional theory. Representative 
phase diagrams and order parameter profiles are presented in Sec. III. Our results 
are summarized in Sec. VI.

\section{Model}
In this section we outline some points of the basic description of our 
model of ionic liquid crystals, and we record some details about the 
density functional theory.

\subsection{Intermolecular pair potential}

\begin{figure}[ht!]
\begin{center}
\includegraphics[width=7.5cm, clip]{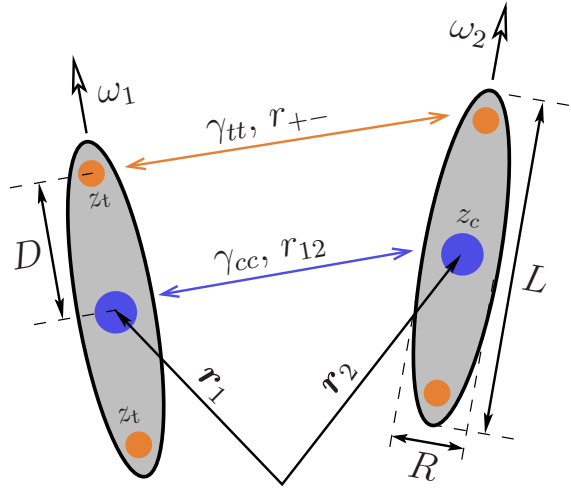}
\caption{Schematic side view of two prolate ellipsoids with orientations 
$\mbox{\boldmath$\omega$}_1$ and $\mbox{\boldmath$\omega$}_2$. The centers of mass 
of the ellipsoids are located at ${\bf r}_1$ and ${\bf  r}_2$, respectively, 
where $r_{12}=|{\bf r}_1 - {\bf  r}_2|$. Only the projections of the ellipsoids on 
the plane of the figure are shown and the centers and main symmetry axes of the 
ellipsoids are chosen to lie within the plane of the figure. $R$ is the cross-sectional 
diameter of the ellipsoids and $L$ is the particle length. The full circles mark the 
location of possible charges with valencies $z_c$ and $z_t$ in the center of the 
ellipsoids and at the tails at a distance $D$ from the center, respectively. The 
energy scale of the Coulomb pair interaction acting between charged tails and between 
charged centers are denoted as $\gamma_{tt}$ and $\gamma_{cc}$ (see Eqs.~(\ref{eq6}) 
and (\ref{eq8})). In addition there is a corresponding Coulomb pair interaction 
acting between charged tails and centers.}
\label{fig1}
\end{center} 
\end{figure}

The intermolecular pair potential is expressed as a sum of the contribution 
due to excluded-volume interactions and the contribution due to long-ranged 
interactions:
\begin{eqnarray} \label{eq1}
U({\bf r}_{12},\mbox{\boldmath$\omega$}_1,\mbox{\boldmath$\omega$}_2)=
\left\{ \begin{array}{cc} 
\infty\,, 
   & r_{12} < 
R \,\sigma({\bf \hat r}_{12},\mbox{\boldmath$\omega$}_1,
\mbox{\boldmath$\omega$}_2) \\
U_{GB}({\bf r}_{12},\mbox{\boldmath$\omega$}_1,\mbox{\boldmath$\omega$}_2)+
U_{CO}({\bf r}_{12},\mbox{\boldmath$\omega$}_1,\mbox{\boldmath$\omega$}_2)\,, 
   & r_{12} \ge 
R\, \sigma({\bf \hat r}_{12},\mbox{\boldmath$\omega$}_1,\mbox{\boldmath$\omega$}_2) 
\end{array} 
\right.\,\,.
\end{eqnarray}
Here the pair potential $U({\bf r}_{12},\mbox{\boldmath$\omega$}_1,
\mbox{\boldmath$\omega$}_2)$ between particles 1 and 2 
is written as a function of the intermolecular vector ${\bf r}_{12}$ between 
the centers of mass of the two particles, and their orientations 
$\mbox{\boldmath$\omega$}_1$ and $\mbox{\boldmath$\omega$}_2$, where 
$r_{12}=|{\bf r}_{12}|$ is the magnitude of ${\bf r}_{12}={\bf r}_2-{\bf r}_1$ 
(see Fig.~\ref{fig1}). The contact distance 
$R\, \sigma({\bf\hat r}_{12},\mbox{\boldmath$\omega$}_1,\mbox{\boldmath$\omega$}_2)$
depends on the orientations of both particles and on the unit vector 
${\bf \hat r}_{12}={\bf r}_{12}/r_{12}$ between their centers.
%
%
%
%
%

We use the well-known Gay-Berne pair potential as a generalization of the 
Lennard-Jones pair potential to fluids consisting of nonspherical 
particles 
(see, e.g., Refs.\cite{gay:81,migu:91,migu:96,brow:98,bate:99,migu:02} and 
references therein):
\begin{eqnarray}
U_{GB}({\bf r}_{12},\mbox{\boldmath$\omega$}_1,\mbox{\boldmath$\omega$}_2)&=&
4 \epsilon({\bf\hat r}_{12},\mbox{\boldmath$\omega$}_1,\mbox{\boldmath$\omega$}_2)\nonumber
\\&\times&
\left[\left(\frac{r_{12}}{R_0}-
\sigma({\bf\hat r}_{12},\mbox{\boldmath$\omega$}_1,\mbox{\boldmath$\omega$}_2)+1\right)^{-12}
-\left(\frac{r_{12}}{R_0}-\sigma({\bf\hat r}_{12},\mbox{\boldmath$\omega$}_1,
\mbox{\boldmath$\omega$}_2)+1\right)^{-6}\right],                       \label{eq2}
\\\sigma({\bf\hat r}_{12},\mbox{\boldmath$\omega$}_1,\mbox{\boldmath$\omega$}_2)&=&
\left[
1-\frac{\chi}{2}\left(
\frac{({\bf \hat r}_{12}\cdot \mbox{\boldmath$\omega$}_1+
{\bf \hat r}_{12}\cdot \mbox{\boldmath$\omega$}_2)^2}
{1+\chi \mbox{\boldmath$\omega$}_1 \cdot  \mbox{\boldmath$\omega$}_2}+
\frac{({\bf \hat r}_{12}\cdot \mbox{\boldmath$\omega$}_1-
{\bf \hat r}_{12}\cdot \mbox{\boldmath$\omega$}_2)^2}
{1-\chi \mbox{\boldmath$\omega$}_1 \cdot  \mbox{\boldmath$\omega$}_2}
\right)\right]^{-\frac{1}{2}}   \,,                                           \label{eq3}
\\\epsilon({\bf\hat r}_{12},\mbox{\boldmath$\omega$}_1,\mbox{\boldmath$\omega$}_2)&=&
\epsilon_0 \left(1-\chi^2 (\mbox{\boldmath$\omega$}_1 \cdot 
 \mbox{\boldmath$\omega$}_2)^2\right)^{-\frac{1}{2}}\nonumber
\\&\times&
\left[1-\frac{\chi'}{2}\left(
\frac{({\bf \hat r}_{12}\cdot \mbox{\boldmath$\omega$}_1+
{\bf \hat r}_{12}\cdot \mbox{\boldmath$\omega$}_2)^2}
{1+\chi' \mbox{\boldmath$\omega$}_1 \cdot  \mbox{\boldmath$\omega$}_2}+
\frac{({\bf \hat r}_{12}\cdot \mbox{\boldmath$\omega$}_1-
{\bf \hat r}_{12}\cdot \mbox{\boldmath$\omega$}_2)^2}
{1-\chi' \mbox{\boldmath$\omega$}_1 \cdot \mbox{\boldmath$\omega$}_2}
\right)\right]^2\,.                                                        \label{eq4}
\end{eqnarray}
Here $\chi=(\kappa^2-1)/(\kappa^2+1)$ and $\kappa=L/R$, where $R$ is the 
cross-sectional diameter of the particle and $L$ is the particle length 
along the main symmetry axis (see Fig.~\ref{fig1}). Accordingly, the parameter 
$\kappa$ is a measure of the length-to-breadth ratio of the particle. The interaction 
strength $\epsilon({\bf\hat r}_{12},\mbox{\boldmath$\omega$}_1,\mbox{\boldmath$\omega$}_2)$
depends on the relative orientations of the particles,  
$\epsilon_0$ is a parameter setting the energy scale of the pair interaction, 
$\chi'=(\kappa'^{1/2}-1)/(\kappa'^{1/2}+1)$, and $\kappa'=\epsilon_R/\epsilon_L$.
Here $\epsilon_R$ is the minimum of the potential for a pair of parallel particles
placed side-by-side 
(${\bf \hat r}_{12}\cdot \mbox{\boldmath$\omega$}_1=
{\bf \hat r}_{12}\cdot \mbox{\boldmath$\omega$}_2=0$)
and $\epsilon_L$ is the minimum for a pair of parallel particles placed end-to-end
(${\bf \hat r}_{12}\cdot \mbox{\boldmath$\omega$}_1=
{\bf \hat r}_{12}\cdot \mbox{\boldmath$\omega$}_2=1$).
The pair interaction potential due to the charges is decomposed into 
three terms
\begin{eqnarray}   \label{eq5}
U_{CO}({\bf r}_{12},\mbox{\boldmath$\omega$}_1,\mbox{\boldmath$\omega$}_2)&=&
U_{cc}({\bf r}_{12})+
U_{ct}({\bf r}_{12},\mbox{\boldmath$\omega$}_1,\mbox{\boldmath$\omega$}_2)+
U_{tt}({\bf r}_{12},\mbox{\boldmath$\omega$}_1,\mbox{\boldmath$\omega$}_2)\,,
\end{eqnarray}
with
\begin{eqnarray}
U_{cc}({\bf r}_{12})&=&\gamma_{cc}\frac{e^{-r_{12}/\lambda_D}}{r_{12}}\,,   \label{eq6}
\\U_{ct}({\bf r}_{12},\mbox{\boldmath$\omega$}_1,\mbox{\boldmath$\omega$}_2)&=&
\gamma_{ct}\left[\frac{e^{-r_{c+}/\lambda_D}}{r_{c+}}+\frac{e^{-r_{+c}/\lambda_D}}{r_{+c}}+
\frac{e^{-r_{c-}/\lambda_D}}{r_{c-}}+\frac{e^{-r_{-c}/\lambda_D}}{r_{-c}}\right]\,,   \label{eq7}
\\U_{tt}({\bf r}_{12},\mbox{\boldmath$\omega$}_1,\mbox{\boldmath$\omega$}_2)&=&
\gamma_{tt}\left[\frac{e^{-r_{++}/\lambda_D}}{r_{++}}+\frac{e^{-r_{+-}/\lambda_D}}{r_{+-}}+
\frac{e^{-r_{-+}/\lambda_D}}{r_{-+}}+\frac{e^{-r_{--}/\lambda_D}}{r_{--}}\right]\,. \label{eq8}
\end{eqnarray}
Here the distances between the charges are given by 
\begin{eqnarray}
r_{c \pm}&=&\left|{\bf r}_{12}     \mp \mbox{\boldmath$\omega$}_2 D\right|\,,
\hspace{0.5cm}
r_{\pm c}=\left|{\bf r}_{12}        \pm \mbox{\boldmath$\omega$}_1 D\right|\,,
\hspace{0.5cm}
r_{\pm \pm}=\left|{\bf r}_{12} \pm \mbox{\boldmath$\omega$}_1 D \mp
\mbox{\boldmath$\omega$}_2 D\right|\,,                                          \label{eq9}
\end{eqnarray}
where $D$ is distance between the center of the particle and the charges
at the tails of the particle (see Fig.~\ref{fig1}).
The Debye screening length is denoted as $\lambda_D$ and 
$\gamma_{cc}=z_c z_c e^2/\varepsilon$, $\gamma_{ct}=z_c z_t e^2/\varepsilon$, 
$\gamma_{tt}=z_t z_t e^2/\varepsilon$ characterize the energy scale. Here 
the sites at the center and the tails of the particle carry the charges 
$z_c e$ and $z_t e$, respectively. The permittivity is denoted as $\varepsilon$. 
Any counterions will be considered at the linear response level, e.g., they 
will screen the electrostatic potential on a scale given by the Debye
screening length.

\subsection{Density functional theory}

The number density of the center of mass of a particle at a point ${\bf r}$
with an orientation $\mbox{\boldmath$\omega$}$ is written as 
$\rho({\bf r},\mbox{\boldmath$\omega$})=\rho\, f({\bf r},\mbox{\boldmath$\omega$})$,
where $f({\bf r},\mbox{\boldmath$\omega$})$ represents a dimensionless
distribution function and $\rho=N/V$ is the total number density. Here $N$ is the 
number of particles and $V$ is volume. The equilibrium density profile minimizes 
the grand potential functional
(see, e.g., Refs.\cite{evan:92,harn:05,harn:06,wu:06,harn:08} and references therein):
\begin{eqnarray}       \label{eq10}
\Omega[\{f({\bf r},\mbox{\boldmath$\omega$})\},\rho,T,\mu]&=&
k_BT\rho \left(\ln(4\pi\Lambda^3 \rho)-1\right)V+
F[\{f({\bf r},\mbox{\boldmath$\omega$})\},\rho,T]-\mu \rho V\,,
\end{eqnarray}
where $\mu$ is the chemical potential and $\Lambda$ is the thermal de Broglie 
wavelength. The excess (over the ideal gas) free energy functional 
$F[\{f({\bf r},\mbox{\boldmath$\omega$})\},\rho,T]$ is in general a very 
complicated, highly non-trivial object, because it is a characterizing property 
of a many-body problem. $F[\{f({\bf r},\mbox{\boldmath$\omega$})\},\rho,T]$ 
is dealt with in various ways, which specify the explicit forms of the theory.
We use the Parsons and Lee approach \cite{pars:79,lee:87} for the  hard core 
interaction together with a perturbation expansion for the long-ranged interaction:
\begin{eqnarray}
F[\{f({\bf r},\mbox{\boldmath$\omega$})\},\rho,T]&=&k_BT\rho
\int d{\bf r}_1\, d \mbox{\boldmath$\omega$}_1\, f({\bf r}_1,\mbox{\boldmath$\omega$}_1)
\ln\left(f({\bf r}_1,\mbox{\boldmath$\omega$}_1)\right)\nonumber
\\&+&\frac{\rho}{2}\int d{\bf r}_1\, d \mbox{\boldmath$\omega$}_1\, 
f({\bf r}_1,\mbox{\boldmath$\omega$}_1)
\left(U_{ref}[\{f({\bf r},\mbox{\boldmath$\omega$})\},\rho,T]+
U_{exc}[\{f({\bf r},\mbox{\boldmath$\omega$})\},\rho,T]\right)\,,\nonumber
\\      \label{eq11}
\end{eqnarray}
with 
\begin{eqnarray}
U_{ref}[\{f({\bf r},\mbox{\boldmath$\omega$})\},\rho,T]&=&
-k_BT J(\rho)\int d{\bf r}_2\, d \mbox{\boldmath$\omega$}_2\, 
f_M({\bf r}_{12},\mbox{\boldmath$\omega$}_1,\mbox{\boldmath$\omega$}_2)
f({\bf r}_2,\mbox{\boldmath$\omega$}_2)\,,                             \label{eq12}
\\U_{exc}[\{f({\bf r},\mbox{\boldmath$\omega$})\},\rho,T]&=&
\rho\int d{\bf r}_2\, d \mbox{\boldmath$\omega$}_2\, 
\left(1-f_M({\bf r}_{12},\mbox{\boldmath$\omega$}_1,\mbox{\boldmath$\omega$}_2)\right)
U({\bf r}_{12},\mbox{\boldmath$\omega$}_1,\mbox{\boldmath$\omega$}_2)
f({\bf r}_2,\mbox{\boldmath$\omega$}_2)\,.                             \label{eq13}
\end{eqnarray}
Here $f_M({\bf r}_{12},\mbox{\boldmath$\omega$}_1,\mbox{\boldmath$\omega$}_2)$ 
is the Mayer function of the hard core pair interaction potential between 
two particles. The Mayer function equals -1 if the particles overlap, i.e., 
$r_{12} < R \,\sigma({\bf \hat r}_{12},\mbox{\boldmath$\omega$}_1,
\mbox{\boldmath$\omega$}_2)$, and is zero otherwise. 
We note that the range parameter
$R\, \sigma({\bf \hat r}_{12},\mbox{\boldmath$\omega$}_1,\mbox{\boldmath$\omega$}_2)$
given by Eq.~(\ref{eq3}) is, to a first approximation, the contact distance between 
two hard ellipsoids of elongation $\kappa$ with orientations 
$\mbox{\boldmath$\omega$}_1$ and $\mbox{\boldmath$\omega$}_2$ 
(see Ref.~\cite{clea:96}).
$U_{ref}[\{f({\bf r},\mbox{\boldmath$\omega$})\},\rho,T]$ and
$U_{exc}[\{f({\bf r},\mbox{\boldmath$\omega$})\},\rho,T]$ are the effective 
reference and excess potential, respectively, acting on a particle due to the 
presence of the remaining particles. The Parsons-Lee modification amounts to 
scaling the reference Onsager free energy functional by the function 
(see Eq.~(\ref{eq12}))

\begin{eqnarray} \label{eq14}
J(\rho)=\frac{1}{V_p}\frac{4\eta-3\eta^2}{4(1-\eta)^2}
\end{eqnarray}
which incorporates the contributions of many-body hard core interactions 
in an approximate way. Here $\eta=\rho V_p$ and $V_p=\pi L R^2/6$ are the 
volume fraction and the particle volume, respectively. In the case of 
$J(\rho)=\rho$ the effective reference potential 
$U_{ref}[\{f({\bf r},\mbox{\boldmath$\omega$})\},\rho,T]$ 
reduces to the original second-virial Onsager theory. The Parsons-Lee 
approach for thermodynamic properties of the isotropic and nematic 
phases of fluids consisting of hard ellipsoids or spherocylinders 
has been found to be in agreement with simulation data.
\cite{samb:94,camp:96,grot:96,melg:08}

The equilibrium distribution function $f({\bf r},\mbox{\boldmath$\omega$})$ 
is obtained from the extremum condition through the variation of the 
excess free energy functional with respect to $f({\bf r},\mbox{\boldmath$\omega$})$:
\begin{eqnarray}     \label{eq15}
f({\bf r},\mbox{\boldmath$\omega$})=\frac{1}{Z[\{f({\bf r},\mbox{\boldmath$\omega$})\},\rho,T]}
\exp\left(-\frac{U_{ref}[\{f({\bf r},\mbox{\boldmath$\omega$})\},\rho,T]}{k_BT}
-\frac{U_{exc}[\{f({\bf r},\mbox{\boldmath$\omega$})\},\rho,T]}{k_BT}\right)\,,
\end{eqnarray}
where the partition function is given by 
\begin{eqnarray}    \label{eq16}
Z[\{f({\bf r},\mbox{\boldmath$\omega$})\},\rho,T]=\frac{1}{V}
\int d{\bf r}\, d \mbox{\boldmath$\omega$}\, 
\exp\left(-\frac{U_{ref}[\{f({\bf r},\mbox{\boldmath$\omega$})\},\rho,T]}{k_BT}
-\frac{U_{exc}[\{f({\bf r},\mbox{\boldmath$\omega$})\},\rho,T]}{k_BT}\right).
\end{eqnarray}
By substituting the distribution function $f({\bf r},\mbox{\boldmath$\omega$})$ back 
into the Eq.~(\ref{eq11}) the excess free energy functional can be written as
\begin{eqnarray}    \label{eq17}
F[\{f({\bf r},\mbox{\boldmath$\omega$})\},\rho,T] V^{-1}&=&
-k_BT\rho \ln\left(Z[\{f({\bf r},\mbox{\boldmath$\omega$})\},\rho,T]\right)
-\frac{\rho}{2}\left\langle U_{ref}[\{f({\bf r},\mbox{\boldmath$\omega$})\},\rho,T]\right\rangle
\nonumber
\\&-&\frac{\rho}{2}\left\langle U_{exc}[\{f({\bf r},\mbox{\boldmath$\omega$})\},\rho,T]\right\rangle\,,
\end{eqnarray}
where the average of a general functional of the distribution function 
$A[\{f({\bf r},\mbox{\boldmath$\omega$})\}]$ is
\begin{eqnarray}    \label{eq18}
\left\langle A[\{f({\bf r},\mbox{\boldmath$\omega$})\}] \right\rangle=\frac{1}{V}
\int d{\bf r}\, d \mbox{\boldmath$\omega$}f({\bf r},\mbox{\boldmath$\omega$})
 A[\{f({\bf r},\mbox{\boldmath$\omega$})\}]\,.
\end{eqnarray}
Minimization of $\Omega[\{f({\bf r},\mbox{\boldmath$\omega$})\},\rho,T,\mu]$ with 
respect to $\rho$ leads to the following Euler-Lagrange equation for the density
\begin{eqnarray}    \label{eq19}
\rho=\frac{e^{\mu/(k_BT)}Z[\{f({\bf r},\mbox{\boldmath$\omega$})\},\rho,T]}
{4\pi \Lambda^3}
\exp\left(-\frac{\left\langle 
U_{ref}[\{f({\bf r},\mbox{\boldmath$\omega$})\},\rho,T]\right\rangle}
{2J(\rho) k_BT}\left(\rho\partial_\rho J(\rho)-J(\rho)\right)\right)\,.
\end{eqnarray}
This equation can be solved numerically for a given chemical potential $\mu$.
The equation of state derived form the grand potential functional takes the 
following form:
\begin{eqnarray}    \label{eq20}
P&=&-\Omega[\{f({\bf r},\mbox{\boldmath$\omega$})\},\rho,T,\mu] V^{-1}
\\&=&
\rho k_BT+\frac{\rho}{2}\left(
\left\langle U_{exc}[\{f({\bf r},\mbox{\boldmath$\omega$})\},\rho,T]\right\rangle
+\frac{\left\langle U_{ref}[\{f({\bf r},\mbox{\boldmath$\omega$})\},\rho,T]\right\rangle}
{J(\rho)}\rho\partial_\rho J(\rho)\right)\,. \label{eq21}
\end{eqnarray}  
The densities, distribution functions, and thermodynamic properties of two coexisting 
phases I and II are found by solving the coexistence conditions $\mu_I=\mu_{II}$ 
and $P_I=P_{II}$ for a given temperature. Here $\mu_I, \mu_{II}$  and 
$P_I, P_{II}$ are the chemical potentials and the pressures of the coexisting 
phases, respectively.

\subsection{Harmonic expansion}

The effective reference and excess potentials are assumed to be of the general form
\begin{eqnarray}   
U_{ref}[S_2,W_0,W_2,d,\rho,T]&=&\frac{J(\rho)}{2}\left(
w_{00}^{(ref)}(\infty,T)+w_{22}^{(ref)}(\infty,T)
P_2(\cos\theta) S_2 \right)\nonumber
\\&+&J(\rho)\cos\left (2\pi z d^{-1} \right)\left(
w_{00}^{(ref)}(d,T) W_0  \right.\nonumber
\\&+&\left.
w_{02}^{(ref)}(d,T) \left(W_2 + P_2(\cos\theta) W_0 \right)
+w_{22}^{(ref)}(d,T)
P_2(\cos\theta) W_2 \right)\,,\nonumber
\\&&                                                  \label{eq22}
\\U_{exc}[S_2,W_0,W_2,d,\rho,T]&=&\frac{\rho}{2}\left(
w_{00}^{(exc)}(\infty)+w_{22}^{(exc)}(\infty)
P_2(\cos\theta) S_2 \right)\nonumber
\\&+&\rho \cos\left( 2\pi z d^{-1} \right)\left(
w_{00}^{(exc)}(d) W_0 + w_{02}^{(exc)}(d) \left(
W_2 + P_2(\cos\theta) W_0 \right)  \right.\nonumber
\\&+&\left.+ w_{22}^{(exc)}(d)
P_2(\cos\theta) W_2 \right) \,,                       \label{eq23}
\end{eqnarray}
where $P_2(\cos\theta)=(3\cos^2\theta-1)/2$ is the second Legendre polynomial 
and $d$ is the layer spacing along $z$ axis in the case of a smectic A phase.
The order parameters $S_2$, $W_0$, and $W_2$ are given by 
\begin{eqnarray}   \label{eq24}
S_2&=&\left\langle P_2(\cos\theta) \right\rangle\,,
\hspace{0.5cm}
W_0=\left\langle \cos\left(2\pi z d^{-1}   \right) \right\rangle\,, 
\hspace{0.5cm}
W_2=\left\langle  P_2(\cos\theta)\cos\left(2\pi z d^{-1}\right) \right\rangle\,.  
\end{eqnarray}
The expansion coefficients $w_{ln}^{(ref)}(d,T)$ and $w_{ln}^{(exc)}(d)$ with 
$ln \in \{00, 02, 22\}$ are evaluated numerically according to 
\begin{eqnarray} 
w_{ln}^{(ref)}(d,T)&=&k_BT\int\limits^\pi_0 d\theta_1\, \sin\theta_1
\int\limits^\pi_0 d\theta_2\, \sin\theta_2 \int\limits^{2\pi}_0 d\phi\,
\int d {\bf r}_{12}\, 
\Theta\left(R\, \sigma({\bf r}_{12},\mbox{\boldmath$\omega$}_1,
\mbox{\boldmath$\omega$}_2) - {\bf r}_{12}\right) \nonumber
\\&\times&
\cos\left(2\pi z_{12} d^{-1}\right)
Q_{ln}(\mbox{\boldmath$\omega$}_1,\mbox{\boldmath$\omega$}_2)\,,             \label{eq25}
\\w_{ln}^{(exc)}(d)&=&\int\limits^\pi_0 d\theta_1\, \sin\theta_1
\int\limits^\pi_0 d\theta_2\, \sin\theta_2 \int\limits^{2\pi}_0 d\phi\,
\int d {\bf r}_{12}\, 
\Theta\left({\bf r}_{12} - R\, \sigma({\bf r}_{12},
\mbox{\boldmath$\omega$}_1,\mbox{\boldmath$\omega$}_2)\right) \nonumber
\\&\times&
U({\bf r}_{12},\mbox{\boldmath$\omega$}_1,\mbox{\boldmath$\omega$}_2)
\cos\left(2\pi z_{12} d^{-1}\right)
Q_{ln}(\mbox{\boldmath$\omega$}_1,\mbox{\boldmath$\omega$}_2) \nonumber
\\               \label{eq26}
\end{eqnarray}
with
\begin{eqnarray} 
Q_{00}(\mbox{\boldmath$\omega$}_1,\mbox{\boldmath$\omega$}_2)&=&
\frac{1}{4\pi}\,,  
\hspace*{0.5cm}
Q_{02}(\mbox{\boldmath$\omega$}_1,\mbox{\boldmath$\omega$}_2)=
\frac{5}{4\pi}\left(\frac{3}{2}\cos^2\theta_1-\frac{1}{2}\right)\,, 
          \label{eq29}
\\Q_{22}(\mbox{\boldmath$\omega$}_1,\mbox{\boldmath$\omega$}_2)&=&
\frac{25}{4\pi}\left(\frac{3}{2}\cos^2\theta_1-\frac{1}{2}\right)
\left(\frac{3}{2}\cos^2\theta_2-\frac{1}{2}\right)\,.    
          \label{eq30}
\end{eqnarray}
Here $\Theta(z)$ is the Heaviside step function and ${\bf r}_{12}=(x_{12},y_{12},z_{12})$.
Equations (\ref{eq22}) and (\ref{eq23}) represent the first terms of an 
expansion of the effective reference and excess potentials in terms of 
spherical invariants (see, e.g., Refs.~\cite{gork:07a,gork:07b} and 
references therein).

The order parameters $S_2$, $W_0$, and $W_2$ in Eq.~(\ref{eq24}) serve to 
distinguish isotropic ($S_2=W_0=W_2=0$), nematic ($S_2\not=0$ and $W_0=W_2=0$), 
and smectic A ($S_2,W_0,W_2\not=0$) phases. By introducing further order 
parameters it is possible to describe other liquid 
crystalline and crystalline structures. However, some of these additional 
phases are strongly non-uniform and they occur at high packing fractions, 
such that they are not expected to be well described by our approach and we 
therefore restrict ourselves to the three order parameters given in Eq.~(\ref{eq24}).

\section{Results}
In this section we discuss fluid phase equilibria for nonspherical particles
with the intermolecular pair potential given by Eq.~(\ref{eq1}). The phase 
diagrams have been calculated using the formalism presented in the previous 
section. In particular, the harmonic expansions of the effective reference and 
excess potentials [Eqs.~(\ref{eq22}) and (\ref{eq23})] have been used as input 
into the expressions for the equilibrium distribution function [Eq.~(\ref{eq15})]
and the pressure [Eq.~(\ref{eq21})]. It is convenient to examine the phase 
behavior in terms of the reduced temperature $T^\star=k_BT/\epsilon_0$, the 
packing fraction $\eta=\rho V_p$, the dimensionless strengths of the Coulomb
interactions $E_{cc}=\gamma_{cc}/(\epsilon_0 R)$, 
$E_{ct}=\gamma_{ct}/(\epsilon_0 R)$, $E_{tt}=\gamma_{tt}/(\epsilon_0 R)$, and 
the reduced Debye screening length $\lambda_D^\star=\lambda_D/R$.

\subsection{Influence of Gay-Berne potential and charges on the phase behavior}
\begin{figure}[ht!]
\begin{center}
\includegraphics[width=6.5cm, clip]{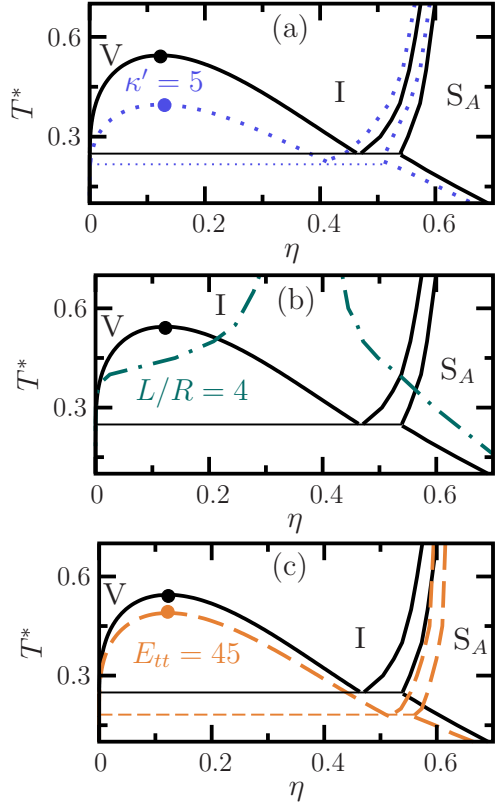}
\caption{The influence of the anisotropy parameter of the Gay-Berne potential 
$\kappa'$ in panel (a), the length-to-breadth ratio of the particles $\kappa=L/R$ 
in panel (b), and like charges at the tails of the particles in panel (c) on the 
fluid phase behavior of an 
ionic liquid crystal consisting of ellipsoidal particles (see Fig.~\ref{fig1}).
The phase diagrams are shown as functions of the packing fraction $\eta$ and the 
reduced temperature $T^\star$. The solid lines in panels (a), (b), and (c) represent the 
phase diagram for uncharged particles with $\kappa=2$ and  $\kappa'=2$, while the 
dotted and dash-dotted lines show the phase diagram for uncharged particles with 
$\kappa=2$, $\kappa'=5$ in panel (a) and $\kappa=4$, $\kappa'=2$ in panel (b). The 
dashed line in panel (c) represents the phase diagram for charged particles with
$\kappa=2$, $\kappa'=2$, $D/R=0.9$, $E_{tt}=45$, and $\lambda^\star_D=50$. The 
solid horizontal lines in panels (a), (b), and (c) mark the coexistence of a vapor 
phase (V) with an isotropic liquid phase (I) and a smectic phase A ($S_A$) for the 
uncharged particles with $\kappa=2$ and  $\kappa'=2$, while the solid circles mark 
the vapor-liquid critical point. The lower circles in panels (a) and (c) denote 
the vapor-liquid critical points corresponding to the phase diagrams represented by 
the dotted and dashed lines, respectively. Vapor-liquid coexistence is metastable for 
the length-to-breadth ratio $\kappa=4$ in panel (b). The dashed horizontal line in 
panel (c) corresponds to V-I-$S_A$ three phase coexistence of the 
fluid consisting of charged particles.}
\label{fig2}
\end{center} 
\end{figure}

First we study the phase behavior of uncharged particles with the 
length-to-breadth ratio $\kappa=L/R=2$ and the anisotropy parameter 
$\kappa'=2$ of the Gay-Berne potential (solid lines in 
Figs.~\ref{fig2} (a) - (c)). The fluid 
is positionally and orientationally disordered ($S_2=W_0=W_2=0$) in the 
isotropic phase (I) at low packing fractions $\eta$ and high enough 
temperatures $T^\star$. Upon increasing the packing fraction, a first-order 
phase transition to a smectic A phase (S$_A$) with $S_2, W_0, W_2 \neq 0$ occurs. 
The isotropic fluid undergoes a vapor-liquid separation below the critical 
temperature $T_c^\star$ marked by the solid circles in Figs.~\ref{fig2} (a) - (c). 
Upon increasing the packing fraction, the phase sequence is vapor (V),
isotropic liquid, and smectic A for temperatures $T_t^\star<T^\star<T_c^\star$. 
Here $T_t^\star$ is the triple point temperature (thin solid line in 
Figs.~\ref{fig2} (a) - (c)) at which the three phases V, I, and $S_A$ coexist.
Increasing the anisotropy parameter $\kappa'$ of the Gay-Berne potential 
at fixed $\kappa$ leads to a shift of the vapor-liquid coexistence curve to 
lower temperatures as is apparent from Fig.~\ref{fig2} (a) where the phase 
diagram is shown for $\kappa'=2$ (solid line) and $\kappa'=5$ (dotted line). 
Moreover, the smectic region is pushed to lower packing fractions as $\kappa'$
increases. High values of $\kappa'$ favor the side-by-side configuration 
over the end-to-end configuration of two parallel particles. Therefore, 
the packing fractions of the coexisting isotropic and smectic A phases 
decrease upon increasing $\kappa'$. The relative 
stability of the side-by-side configuration decreases as $\kappa'$ is lowered 
and for $\kappa'=1$ all configurations are equally stable for parallel 
particles, i.e., 
$\epsilon({\bf r}_{12},\mbox{\boldmath$\omega$}_1,\mbox{\boldmath$\omega$}_2)
=\epsilon_0$ for $\kappa'=1$ and $\mbox{\boldmath$\omega$}_1 \parallel 
\mbox{\boldmath$\omega$}_2$ in Eq.~(\ref{eq4}). 
%
%
%
%

Figure  \ref{fig2} (b) demonstrates that the width of the I-$S_A$ phase
transition broadens upon increasing the length-to-breadth ratio of the particles 
from $\kappa=L/R=2$ (solid line) to $\kappa=4$ (dash-dotted line). Furthermore, the 
vapor-liquid coexistence curve is metastable with respect to the I-$S_A$ coexistence 
for $\kappa=4$. More details concerning the influence of the length-to-breadth 
ratio of the particles and the anisotropy parameter of the Gay-Berne potential 
on the fluid phase behavior can be found in Refs.~\cite{migu:96,brow:98}.

We examine now the influence of two like charges ($E_{tt}=45, \lambda^\star_D=50$)
located at the tails ($D/R=0.9$ for $\kappa=D/R=2$, see Fig.~\ref{fig1}) on the 
fluid phase behavior (dashed line in Fig.~\ref{fig2} (c)). The vapor-liquid 
critical temperature is seen to decrease with increasing the Coulomb 
interaction strength and the I-$S_A$ coexistence region is shifted to higher packing 
fractions. In the high-temperature limit the thermodynamic properties of the 
fluid are dominated by the repulsive steric interactions, and the I-$S_A$ phase 
transition tends to that of the corresponding hard core fluid, with packing 
fractions $\eta_I=0.61$ and $\eta_{S_A}=0.63$ at I-$S_A$ phase coexistence.
The decrease of the vapor-liquid critical temperature is due to the repulsive 
Coulomb pair interaction between the like charged tails of the particles.

\subsection{Influence of the location of charges on the phase behavior}
\begin{figure}[ht!]
\begin{center}
\includegraphics[width=6.5cm, clip]{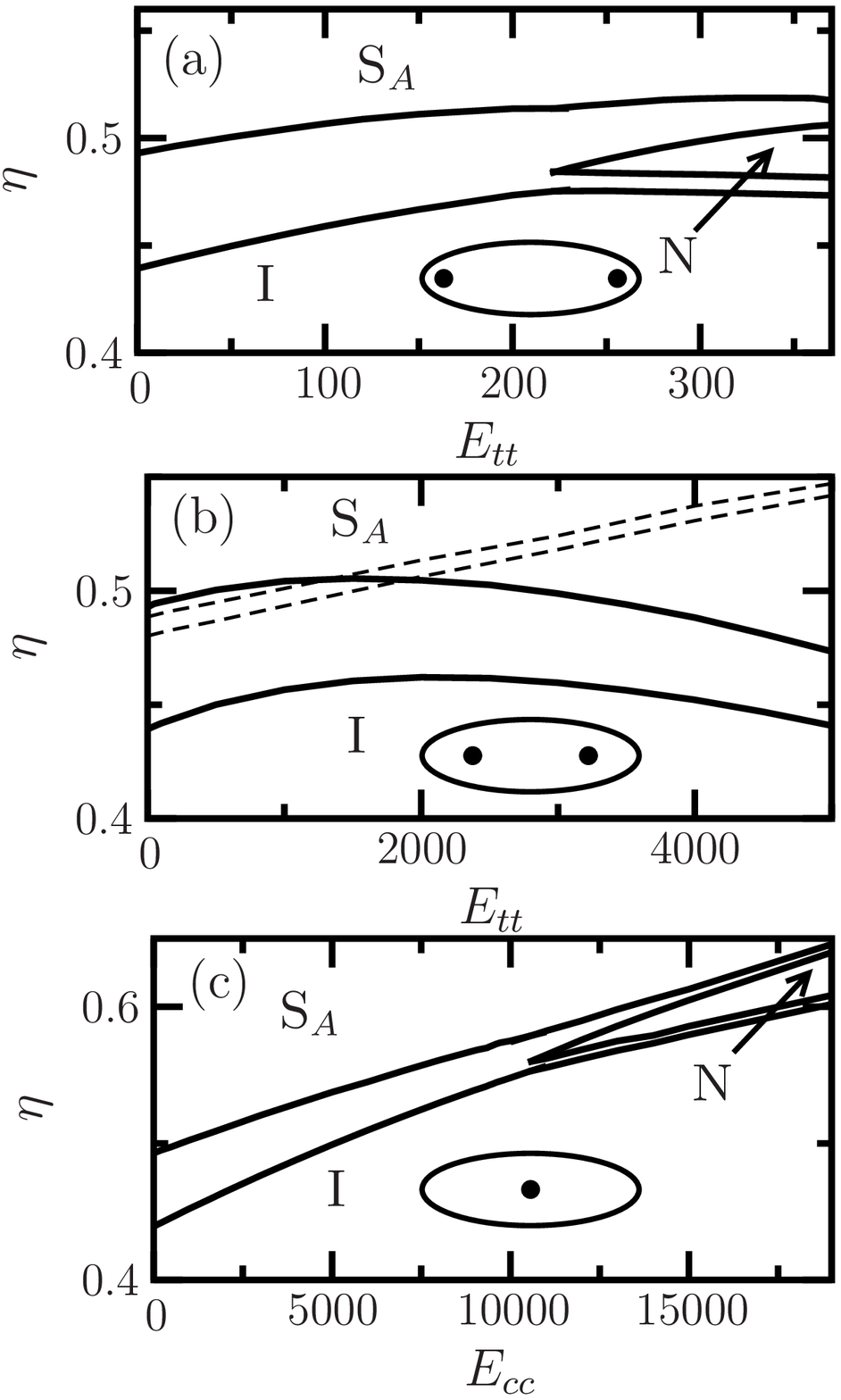}
\caption{The influence of the location of two like charges on the fluid 
phase behavior of an ionic liquid crystal consisting of charged ellipsoidal 
particles with length-to-breadth ratio $\kappa=L/R=3$ at temperature $T^\star=1.5$. 
The anisotropy parameter of the Gay-Berne potential and the Debye screening 
length are fixed to $\kappa'=8$ and $\lambda_D^\star=50$, respectively.
Two like charges are located at the distance $D=1.4 R$ and $D=R$ from the center 
of the particles in panels (a) and (b), respectively, while a single charge is 
located at the center of the particles in panel (c). Schematic illustrations of the 
shape of the particles and the location of the charges (solid dots) are shown in the 
figures. The phase diagrams are plotted as functions of the Coulomb pair 
interaction strengths $E_{tt}$ or $E_{cc}$, and the packing fraction $\eta$.
The solid lines denote the phase boundaries of thermal equilibrium of 
an isotropic (I), nematic (N), and smectic A ($S_A$) phase, while the 
dashed lines in panel (b) mark metastable isotropic-nematic phase coexistence.}
\label{fig3}
\end{center} 
\end{figure}

The effect of varying the location of charges on the particles 
with a fixed length-to-breadth ratio $L/R=3$ and at a fixed temperature $T^\star=1.5$ 
is now examined. Fluid phases are shown in Fig.~\ref{fig3} as functions of the strength 
of the Coulomb pair interactions $E_{tt}$ or $E_{cc}$, and the packing fraction $\eta$
for three different locations of the charges. Two like charges 
are located at the distance $D=1.4 R$ and $D=R$ from the center of the particles in 
Fig.~\ref{fig3} (a) and (b), respectively, while a single charge is located at the 
center of the particles in Fig.~\ref{fig3} (c). The length-to-breadth ratio and 
temperature have been chosen such that the fluids consisting of uncharged particles
(i.e., $E_{tt}=E_{cc}=0$) are isotropic at low packing fractions and a phase transition 
to the smectic A phase is observed at higher packing fractions. There is an important 
difference between the phase behavior of fluids consisting of particles with charges 
located at $D=1.4 R$ and at $D=R$.
While for $D=R$ the smectic A phase is the only stable phase at high packing 
fractions (see Fig.~\ref{fig3} (b)), nematic phase ordering (N) with $S_2\neq 0$ and 
$W_0=W_2=0$ is found for $D=1.4 R$ (see Fig.~\ref{fig3} (a)). In this case the 
nematic phase is stable for strong Coulomb pair interaction down to $E_{tt}=215$ 
at the I-N-$S_A$ triple point. In the case of particles with two like charges 
located at the distance $D=R$ from the center, there is no stable nematic phase even 
at higher Coulomb interaction strengths. In this case the packing fractions of the 
coexisting isotropic and smectic A phases (solid lines in Fig.~\ref{fig3} (b)) 
decrease with increasing Coulomb interaction strength $E_{tt}$, whereas the 
packing fractions of the metastable isotropic-nematic phase coexistence 
(dashed lines in Fig.~\ref{fig3} (b)) increase. 
%
%
%
%

Surprisingly, the nematic phase is stable in the case of a fluid 
consisting of particles with a single charge located at the center as is shown 
in Fig.~\ref{fig3} (c). The smectic A phase is preempted by the nematic phase 
which is stable above the Coulomb interaction strength $E_{cc}=10465$ 
at the I-N-$S_A$ triple point.

\begin{figure}[ht!]
\begin{center}
\includegraphics[width=6.5cm, clip]{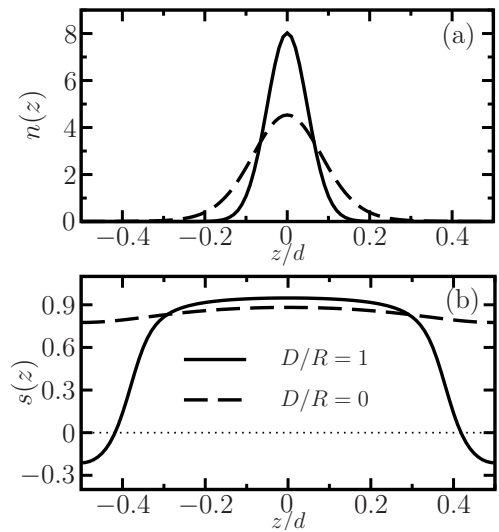}
\caption{Normalized, orientationally averaged density profile $n(z)$ 
[Eq.~(\ref{eq31})] in panel (a) and relative nematic order parameter $s(z)$
[Eq.~(\ref{eq32})] in panel (b) of an ionic liquid crystal along the 
$z$-axis in the smectic A phase, where $d$ is the layer spacing. The solid lines 
represent the profiles for particles with two like charges located at the distance 
$D=R$ from the center of the particles (see Fig.~\ref{fig3} (b)), while the dashed 
lines show the profiles for particles with a single charge located at the center 
of the particles (see Fig.~\ref{fig3} (c)). The model parameters are fixed to 
$\kappa=L/R=3$, $\kappa'=8$, $T^\star=1.5$, $\eta=0.67$, and $E_{tt}=E_{cc}/4=4000$.
Therefore, the particles in panels (a) and (b) carry the same total charge.}
\label{fig4}
\end{center} 
\end{figure}

In order to understand the influence of the location of two like charges on the 
fluid phase behavior it is instructive to consider a set of position-dependent 
order parameters which quantifies the deviation of the number density from 
isotropy. \cite{harn:02} The normalized, orientationally averaged density profile 
\begin{equation} \label{eq31} 
n(z)=2 \pi \int\limits_0^{\pi}d\theta\, \sin\theta\,f(z,\theta) 
\end{equation} 
and the position-dependent, relative nematic order parameter
\begin{equation} \label{eq32} 
s(z)=\frac{\pi }{n(z)} \int\limits_0^{\pi}d\theta\, 
\sin\theta\,(3\cos^2\theta-1) f(z,\theta) 
\end{equation} 
are displayed in Fig.~\ref{fig4} (a) and (b), respectively. The solid lines show 
the profiles for particles with two like charges located at the distance $D=R$ from the 
center of the particles (see Fig.~\ref{fig3} (b)), while the dashed lines display 
the profiles for particles with a single charge located at the center of the particles 
(see Fig.~\ref{fig3} (c)). The packing fraction is fixed to $\eta=0.67$ and the 
Coulomb pair interaction strengths are given by $E_{tt}=E_{cc}/4=4000$. Hence
the smectic A phase is stable and the particles in panels (a) and (b) carry the same 
total charge. The order parameter profiles are periodic functions with layer spacings 
$d \approx 3.8 R$ and $d \approx 3.0 R$ for the particles with two like charges at 
the tails (solid lines) and the particles with a single charge located at the center 
(dashed lines), respectively. 
The density profiles of the centers of the particles exhibit maxima in the center
of the layers at $z=0$ as is apparent from Fig.~\ref{fig4} (a). Moreover, the density
distribution along the layer axis is sharper for the particles with two like charges 
at the tails than that for the particles with a single charge located at the center.
%
%
%
%

Figure \ref{fig4} (b) demonstrates that there is a qualitative difference between 
the relative nematic order parameter profiles for particles with two like charges 
at the tails (solid lines) and the particles with a single charge located at the
center (dashed lines). Whereas $s(z)$ is rather independent of $z$ in the 
latter case, the relative nematic order parameter profile exhibits pronounced 
oscillations along the $z$ axis in the former case, where particles located between 
the layers at $|z/d| \gtrsim 0.4$ are oriented with their main body mainly perpendicular 
to the $z$ axis, i.e., $s(z)<0$. For comparison we recall that the value of the nematic 
order parameter is $s(z)=-0.5$ and $s(z)=1.0$ for perfect perpendicular and parallel 
alignment to the $z$ axis, respectively.
The predominantly perpendicular orientation of particles with charges at the tails
located in between the smectic layers can be understood in terms of a minimization of
the electrostatic repulsion due to a maximized distance from the particles in the smectic
layers. 
For particles with the charge in the center the electrostatic energy is independent of 
the orientation; hence a parallel alignment of particles in between smectic layers is 
favourable, as non-parallel orientations would increase the free energy due to an increase 
in the layer spacing. 
The latter case is comparable with the results of van Roij et al.~\cite{roij:95}
on uncharged spherocylinders (see Fig. 2 of Ref.~\cite{roij:95}), where the majority of inter-layer particles is aligned parallel to the layer normal.
The bimodal orientational distribution described in Ref.~\cite{roij:95} is also 
expected to be found in the present situation.

\subsection{Influence of the particle length on the phase behavior}

\begin{figure}[ht!]
\begin{center}
\includegraphics[width=6.5cm, clip]{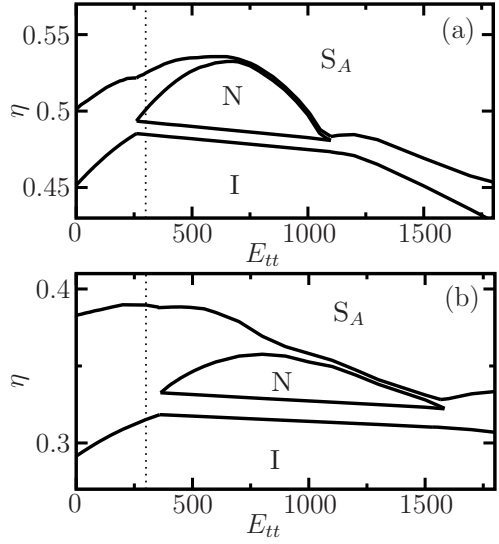}
\caption{The influence of the length-to-breadth ratio $\kappa=L/R$ 
on the fluid phase behavior of an ionic liquid crystal consisting of charged
ellipsoidal particles with two like charges located at a fixed distance 
$L/2-D=0.1 R$ from the end of the particles (see Fig.~\ref{fig1}). 
The length-to-breadth ratio is $\kappa=3$ in panel (a) and $\kappa=5$ in panel 
(b). The anisotropy parameter of the Gay-Berne potential and the Debye 
screening length are fixed to $\kappa'=8$ and $\lambda_D^\star=50$, 
respectively. The phase diagrams are plotted as functions of the Coulomb 
pair interaction strength $E_{tt}$ and the packing fraction $\eta$ for 
the fixed temperature $T^\star=2$. The solid lines denote the phase 
boundaries of thermal equilibrium of an isotropic (I), nematic (N), 
and smectic A ($S_A$) phase. In Fig.~\ref{fig6} phase diagrams 
are shown as functions of the packing fraction and the temperature 
for the Coulomb pair interaction strength indicated by the dotted lines.}
\label{fig5}
\end{center} 
\end{figure}

We now study the effect of varying the length-to-breadth ratio $\kappa=L/R$ for 
charged particles with two like charges located at a fixed distance 
$L/2-D=0.1 R$ from the end of the particles (see Fig.~\ref{fig1}). Fluid phases 
are shown in Fig.~\ref{fig5} as functions of the strength of the Coulomb pair 
interaction $E_{tt}$ and the packing fraction $\eta$ for a fixed 
temperature $T^\star=2$. The length-to-breadth ratio is $\kappa=3$ 
and $\kappa=5$ in Fig.~\ref{fig5} (a) and (b), respectively. Qualitatively 
similar types of phase behavior are exhibited by both systems. The fluids consisting 
of uncharged particles, i.e., $E_{tt}=0$, are isotropic at low packing fractions 
and a phase transition to the smectic A phase is observed at higher packing 
fractions. Upon increasing the Coulomb pair interaction strength $E_{tt}$ 
stable nematic islands in the phase diagrams are found. This nematic phase 
is bounded below and above by isotropic and smectic A phases, respectively.
Moreover, the location of the nematic region is seen to move to higher Coulomb 
pair interaction strength and lower volume fraction upon increasing the 
length-to-breadth ratio. The nematic phase disappears at high values of the 
Coulomb pair interaction strength when the repulsive steric interaction is less 
important. The competition of the steric interaction and the Coulomb pair 
interaction leads to the existence of a stable nematic phase for intermediate 
values of $E_{tt}$. Moreover, we emphasize that the smectic A phase is stabilized 
for high Coulomb interaction strengths. Hence molecules which are not 
mesogenic without charges at a given packing fraction can form a stable 
smectic phase if they are charged at the same packing fraction. Furthermore, 
it is worthwhile to note that using 
larger Debye screening lengths leads to phase diagrams of the same topology 
as the ones presented in Fig.~\ref{fig5} (data not shown). However, the 
corresponding Coulomb interaction strengths are smaller due to the longer 
ranged interaction potential.
%
%
%
%

\begin{figure}[ht!]
\begin{center}
\includegraphics[width=6.5cm, clip]{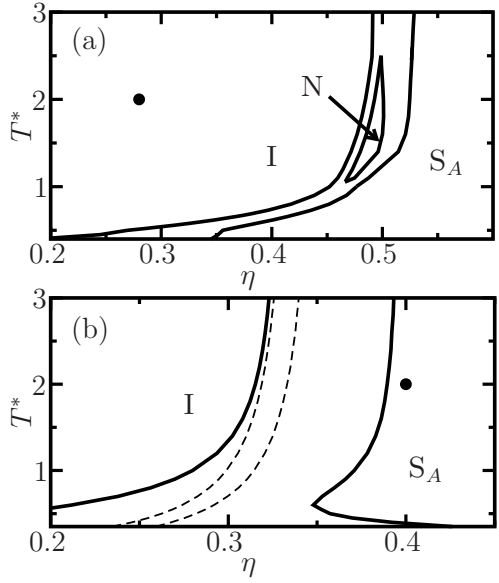}
\caption{Phase diagrams of the same fluids as in Fig.~\ref{fig5} 
in the packing fraction ($\eta$) - temperature ($T^\star$) plane. The 
Coulomb pair interaction strength is given by $E_{tt}=300$ as indicated by 
the dotted lines in Fig.~\ref{fig5}. The length-to-breadth ratio is $\kappa=3$ 
in panel (a) and $\kappa=5$ in panel (b). The solid lines denote the phase 
boundaries of thermal equilibrium of an isotropic (I), nematic (N), and smectic A 
($S_A$) phase, while the dashed lines in panel (b) mark metastable isotropic-nematic
phase coexistence. In panel (a) and (b) the solid circles denote two state 
points with equal temperature and pressure.}
\label{fig6}
\end{center} 
\end{figure}

Packing fraction-temperature projections of the fluid phase diagrams for the 
systems with $E_{tt}=300$ (see the dotted lines in Fig.~\ref{fig5})
are shown in Fig.~\ref{fig6}. The stable isotropic, nematic, and 
smectic A regions are clearly visible for the fluid consisting of the smaller
particles in Fig.~\ref{fig6} (a). For the larger particles the isotropic-nematic
coexistence region is metastable with respect to the I-$S_A$ coexistence 
(dashed lines in Fig.~\ref{fig6} (b)). For both systems, the low temperature 
part of the phase diagram is dominated by a wide two-phase region where the 
$S_A$ phase is in equilibrium with an isotropic phase. We note that the behavior 
is reverse for $E_{tt}^{(l)} > E_{tt} > E_{tt}^{(s)}$, where $E_{tt}^{(s)}=1099$ 
and $E_{tt}^{(l)}=1570$ is the second triple point for the smaller and larger 
particles, respectively (see Fig.~\ref{fig5} (a)). Increasing the
length-to-breadth ratio induces the nematic phase in this case.
%
%
%
%

The model described in Sec.~IIA of ellipsoidal particles with point charges
may be compared with models of spherocylinders with line charges. \cite{graf:99,kram:00,egge:09}
The presence of a direct isotropic-smectic transition for small length-to-breadth
ratios $\kappa=L/R$ and charges $E_{tt}$, the decrease of the packing fraction 
at the transition upon increasing $\kappa$, as well as the increase of the packing 
fraction at the transition upon increasing $E_{tt}$ (see Fig.~\ref{fig5}) are in agreement
with the trends for spherocylinders as displayed in Figs.~2 and 3 of Ref.~\cite{kram:00}.
Hence there is qualitative similarity between our model of point charges at the particle
tails and charged spherocylinders with line charges.

\subsection{Varying the strength of the Coulomb interaction}

\begin{figure}[ht!]
\begin{center}
\includegraphics[width=6.5cm, clip]{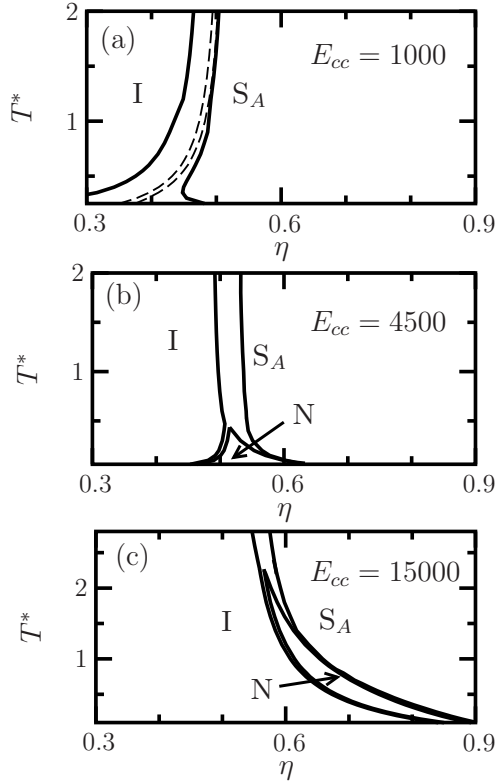}
\caption{The influence of the Coulomb pair interaction strength 
$E_{cc}$ on the fluid phase behavior of an ionic liquid crystal consisting of 
ellipsoidal particles ($\kappa=L/R=3$) with a single charge located in the center 
of each particle. The phase diagrams are shown as functions of the packing fraction 
$\eta$ and the reduced temperature $T^\star$. The solid lines denote the phase 
boundaries of thermal equilibrium of an isotropic (I), nematic (N), and smectic 
A ($S_A$) phase, while the dashed lines in panel (a) mark metastable isotropic-nematic 
phase coexistence. The anisotropy parameter of the Gay-Berne potential and the Debye 
screening length are given by $\kappa'=8$ and $\lambda_D^\star=50$, respectively.}
\label{fig7}
\end{center} 
\end{figure}

We next consider the influence of the Coulomb interaction strength 
on the phase behavior. The locations of the various ordering transitions 
in the case of a fluid consisting of ellipsoidal particles with a single 
charge located in the center of the particles and the length-to-breadth 
$\kappa=3$ are summarized in Fig.~\ref{fig7} for three different Coulomb 
interaction strengths $E_{cc}$. From the phase behavior of the systems 
shown in Fig.~\ref{fig7}, it is apparent that the nematic phase becomes stable 
with increasing Coulomb interaction strength. There are I-N-S$_A$ triple 
points for the Coulomb interaction strengths $E_{cc}=4500$ and 
$E_{cc}=15000$ in Figs.~\ref{fig7} (b) and (c), respectively, while 
the I-N coexistence region is metastable in the case of the weaker Coulomb 
interaction strength considered in Fig.~\ref{fig7} (a). In the high-temperature 
limit the thermodynamic properties of the fluids are dominated by the 
repulsive steric interactions, and the I-$S_A$ phase transition tends to 
that of the corresponding hard core fluid. As expected the isotropic region 
becomes more extensive  as the Coulomb interaction strength is increased. 
In the case of a large Coulomb interaction strength the long-ranged pair 
potential is rather independent of the orientations of the particles because 
the charges are located in the center of the ellipsoids. Therefore the locations 
of phase transitions from the isotropic phase to orientationally ordered phases 
in the low temperature region are shifted to higher packing fractions upon 
increasing the Coulomb interaction strength.
%
%
%
%

\subsection{Comparison with experimental results}

In general the total number density of both thermotropic liquid crystals 
(see, e.g., Ref.~\cite{mcla:64}) and colloidal suspensions consisting 
of charged nonspherical particles (see, e.g., Ref.~\cite{dogi:06}) 
is smaller in the isotropic phase than in the coexisting nematic or smectic 
phase similar to our findings presented 
in Figs.~\ref{fig2}, \ref{fig3}, and \ref{fig5} - \ref{fig7}. Nevertheless, 
it is worthwhile to mention that the density of hard platelike particles 
in a binary mixture of thick and thin platelets can be larger in the 
isotropic phase than in the coexisting nematic phase depending on the 
chemical potentials. \cite{kooi:01a} This remarkable phenomenon of 
isotropic-nematic density inversion has been investigated using a 
two-component density functional theory for hard nonspherical particles. 
\cite{bier:04}

\begin{figure}[ht!]
\begin{center}
\includegraphics[width=6.5cm, clip]{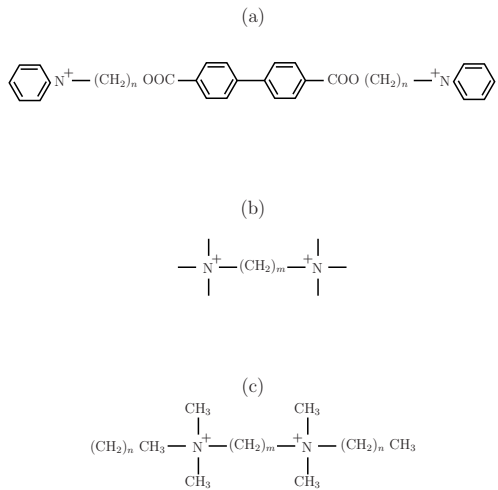}
\caption{Chemical structure of dipolar amphiphiles with two pyridinium head 
groups and a biphenylene core in (a), diquaternary ammonium salts in (b), and 
diammonium salts in (c). These molecules exhibit liquid crystalline phases.
\cite{hess:93,kokk:90,full:96}
}
\label{fig8}
\end{center} 
\end{figure}

Hessel et al. \cite{hess:93} observed isotropic and smectic phases of 
molecules consisting of two pyridinium head groups and a biphenylene core 
(see Fig.~\ref{fig8} (a)). Increasing the length of the alkyl chains, i.e., 
increasing $n$ in Fig.~\ref{fig8} (a), stabilizes the smectic phase. Both the 
transition temperature from the isotropic to the smectic phase and the layer 
spacing in the smectic phase increase with increasing length of the alkyl 
chains. These results agree with the theoretical results presented in 
Figs.~\ref{fig5} and \ref{fig6}. Increasing the length $L$ of particles 
with two like charges located at a fixed distance $L/2-D$ from the center 
of the particles (see Fig.~\ref{fig1}) has a stabilizing effect on the 
smectic phase. For example, the isotropic phase is stable for particles 
with $\kappa=3$ at the state point denoted by the solid circle in 
Fig.~\ref{fig6} (a), while the smectic A phase is s stable for larger 
particles with $\kappa=5$ at the same temperature and pressure marked 
by the solid circle in Fig.~\ref{fig6} (b). Moreover, the calculated layer 
spacing $d$ increases upon increasing $L$.
%
%
%
%

Kokkinia and Paleos \cite{kokk:90} investigated the phase behavior of 
$\alpha\!-\!\omega$ diquaternary ammonium salts (see Fig.~\ref{fig8} (b)).
By increasing the spacer length between the quaternary nitron atoms, 
more stable smectic phases were obtained in agreement with both the 
theoretical results shown in Figs.~\ref{fig5} and \ref{fig6} as well as 
the aforementioned experimental study on dipolar amphiphiles with two 
pyridinium head groups and a biphenylene core. \cite{hess:93}

Thermotropic and lyotropic mesomorphism has been observed for the 
diammonium salts shown in Fig.~\ref{fig8} (c). \cite{full:96}
Interestingly, a stable nematic phase has been found for small spacer 
length $(CH_2)_m$. This nematic phase disappeared upon increasing the 
spacer length similar to our theoretical findings shown in 
Figs.~\ref{fig6} (a) and (b). Moreover, it is worthwhile to mention 
that a decrease of the layer spacing of the smectic phase with increasing 
temperature has been observed for various ionic liquid crystals 
\cite{binn:05} in agreement with the results of the present theoretical 
approach.

\section{Summary and Discussion}

We have investigated bulk properties of ionic liquid crystals using density 
functional theory. The liquid crystal molecules are modeled as charged ellipsoids 
with cross-sectional diameter $R$ and length $L$. Possible charges are located 
in the center of the particles or at the tails at a distance $D$ from the 
center (see Fig.~\ref{fig1}). The intermolecular pair potential is expressed 
as a sum of the contributions due to excluded-volume interactions and the 
contribution due to long-ranged interactions [Eq.~(\ref{eq1})]. These long-ranged 
interactions are taken into account in terms of the Gay-Berne potential 
[Eqs.~(\ref{eq2}) - (\ref{eq4})] and the screened Coulomb interaction 
[Eqs.~(\ref{eq5}) - (\ref{eq9})]. The grand potential functional 
[Eq.~(\ref{eq10})] is minimized numerically and phase diagrams, density profiles, 
and orientational order parameter profiles are determined leading to the 
following main results:

(1) The phase diagrams for representative examples of liquid crystals 
involve a vapor, an isotropic liquid, and a smectic A phase (see 
Fig.~\ref{fig2}). Increasing the anisotropy parameter of the Gay-Berne 
potential leads to a shift of the vapor-liquid coexistence curve to 
lower temperatures (see Fig.~\ref{fig2} (a)). The width of the isotropic 
to smectic phase transition broadens upon increasing the length-to-breadth 
ratio of the particles (see Fig.~\ref{fig2} (b)).

(2) There is a pronounced dependence of the phase behavior on the location 
of two like charges on the ellipsoidal particles (see Fig.~\ref{fig3}). 
While for $D=R$ the smectic A phase is the only stable phase at high packing 
fractions (see Fig.~\ref{fig3} (b)), nematic phase ordering is found for 
$D=1.4 R$ (see Fig.~\ref{fig3} (a)). Moreover, the nematic phase is stable in 
the case of a fluid consisting of particles with a single charge located at 
the center (see Fig.~\ref{fig3} (c)). Whereas the relative nematic order parameter 
profile $s(z)$ is rather independent of $z$ in the smectic A phase of particles 
with a single charge located at the center, it exhibits pronounced oscillations 
along the $z$ axis in the case of two like charges at the tails at a distance 
$D=R$ from the center of the particles (see Fig.~\ref{fig4} (b)).

(3) Increasing the length $L$ of the particles with two like charges located at 
a fixed distance $L/2-D$ from the center of the particles has a stabilizing effect 
on the smectic A phase in agreement with earlier experimental findings (see
Figs.~\ref{fig5} and \ref{fig6}). Moreover, the calculated layer spacing in the 
smectic A phase increases upon increasing the length $L$. With increasing the 
Coulomb pair interaction strength stable nematic islands in the phase diagrams 
are found (see Fig.~\ref{fig5}). This nematic phase is bounded by isotropic and 
smectic A phases. Moreover, the location of the nematic region moves to higher 
Coulomb pair interaction strength and lower volume fraction upon increasing the 
length of the particles.

(4) For particles with a single charge located in the center, the isotropic 
region in the phase diagram becomes more extensive  as the Coulomb interaction 
strength is increased (see Fig.~\ref{fig7}) similar to earlier theoretical 
findings for charged platelike particles. \cite{bier:05} Moreover, a nematic 
phase becomes stable with increasing the Coulomb interaction strength 
(see Figs.~\ref{fig7} (b) and (c)). In the case of a large Coulomb interaction 
strength the long-ranged pair potential is rather independent of the orientations 
of the particles because the charges are located in the center of the ellipsoids.

Finally, we would like to emphasize that the methodology developed here can 
be extended to enable a quantitative treatment of the counterions by 
considering a multi-component density functional theory similar to earlier 
studies of nonspherical particles with charges located in the center.
\cite{bier:05,bier:06} A two-component model can be used to study the influence 
of the small ions on the liquid crystal molecules. On the basis of earlier 
multi-component integral equation studies it is known that the small ions 
screen the Coulomb interaction between the bigger particles in the liquid phase. 
\cite{harn:00,harn:01,harn:02a} This justifies the use of a one-component model 
with the screened Coulomb interaction [Eqs.~(\ref{eq5}) - (\ref{eq9})].
But the two-component model may help to elucidate the influence of possible 
ionic clusters on the liquid-vapor coexistence of room-temperature ionic liquids.
\cite{rebe:05,beta:09}

\section{Acknowledgments}
S. K. and L. H. gratefully acknowledge support by the Deutsche Forschungsgemeinschaft 
under Grant No. HA 2935/4-1. L. H. thanks M. Osipov for useful discussions.


\begin{thebibliography}{99} 

\bibitem{adam:94} D. Adam, P. Schuhmacher, J. Simmerer, L. H\"aussling, 
K. Siemensmeyer, K. H. Etzbach, H. Ringsdorf, and D. Haarer,
Nature {\bf 371}, 141 (1994).

\bibitem{neil:03} M. O'Neill and S. M. Kelly,
Adv. Mater. {\bf 15}, 1135 (2003).

\bibitem{kato:02} T. Kato,
Science {\bf 295}, 2414 (2002).

\bibitem{yosh:02} M. Yoshio, T. Mukai, K. Kanie, M. Yoshizawa, H. Ohno, and T. Kato,
Adv. Mater. {\bf 14}, 351 (2002).

\bibitem{gord:98} C. M. Gordon, J. D. Holbrey, A. R. Kennedy, and K. R. Seddon,
J. Mater. Chem. {\bf 8}, 2627 (1998).

\bibitem{brad:02} A. E. Bradley, C. Hardacre, J. D. Holbrey, 
S. Johnston, S. E. J. McMath, and M. Nieuwenhuyzen, 
Chem. Mater. {\bf 14}, 629 (2002).

\bibitem{lee:03} K.-M. Lee, Y.-T. Lee, and I. J. B. Lin,
J. Mater. Chem. {\bf 13}, 1079 (2003).

\bibitem{roch:03} J. De. Roche., C. M. Gordon, C. T. Imrie, M. D. Ingram, 
A. R. Kennedy, F. Lo Celso, and A. Triolo,
Chem. Mater. {\bf 15}, 3089 (2003).

\bibitem{ster:07} D. Ster, U. Baumeister, J. Lorenzo Chao, C. Tschierske, and G. Israel,
J. Mater. Chem. {\bf 17}, 3393 (2007).

\bibitem{kouw:07} P. H. J. Kouwer and T. M. Swager,
J. Am. Chem. Soc. {\bf 129}, 14042 (2007).

\bibitem{holb:99} J. D. Holbrey and K. R. Seddon,
J. Chem. Soc. Dalton Trans. {\bf 13}, 2133 (1999).

\bibitem{hard:01} C. Hardacre, J. D. Holbrey, P. B. McCormac, 
S. E. J. McMath, M. Nieuwenhuyzen, and K. R. Seddon,
J. Mater. Chem. {\bf 11}, 346 (2001).

\bibitem{saue:09} S. Sauer, S. Saliba, S. Tussetschl\"ager, A. Baro, W. Frey,
F. Giesselmann, S. Laschat, and W. Kantlehner,
Liq. Cryst. {\bf 36}, 275 (2009).

\bibitem{neve:98} F. Neve, A. Crispini, S. Armentano, O. Francescangeli,
Chem. Mater. {\bf 10}, 1904 (1998).

\bibitem{neve:01} F. Neve, O. Francescangeli, A. Crispini, J. Charmant,
Chem. Mater. {\bf 13}, 2032 (2001).

\bibitem{yosh:04a} M. Yoshio, T. Mukai, H. Ohno, and  T. Kato,
J. Am. Chem. Soc. {\bf 126}, 994 (2004).

\bibitem{taub:04} A. Taubert,
Angew. Chem. Int. Ed. {\bf 43}, 5380 (2004).

\bibitem{taub:05} A. Taubert, P. Steiner, and A. Mantion,
J. Phys. Chem. B {\bf 109}, 15542 (2005).

\bibitem{saue:08} S. Sauer, N. Steinke, A. Baro, S. Laschat, F. Giesselmann, 
and W. Kantlehner,
Chem. Mater. {\bf 20}, 1909 (2008).

\bibitem{binn:05} K. Binnemanns,
Chem. Rev. {\bf 105}, 4148 (2005).

\bibitem{gay:81} J. G. Gay and B. J. Berne,
J. Phys. Chem. {\bf 74}, 3316 (1981).

\bibitem{migu:91} E. de Miguel, L. F. Rull, M. K. Chalam, and K. E. Gubbins,
Molec. Phys. {\bf 74}, 405 (1991).

\bibitem{migu:96} E. de Miguel, E. M. del Rio, J. T. Brown, and M. P. Allen,
J. Phys. Chem. {\bf 105}, 4234 (1996).

\bibitem{brow:98} J. T. Brown, M. P. Allen, E. M. del Rio, and E. de Miguel,
Phys. Rev. E {\bf 57}, 6685 (1998).

\bibitem{bate:99} M. A. Bates and G. R. Luckhurst,
J. Phys. Chem. {\bf 110}, 7087 (1999).

\bibitem{migu:02} E. de Miguel and C. Vega,
J. Phys. Chem. {\bf 117}, 6313 (2002).

\bibitem{evan:92} R. Evans in 
{\it Fundamentals of Inhomogeneous Fluids},
edited D. Henderson, p. 85, 
(Dekker, New York, 1992).

\bibitem{harn:05} L. Harnau and S. Dietrich,
Phys. Rev. E {\bf 71}, 011504 (2005).

\bibitem{harn:06} L. Harnau and S. Dietrich in 
{\it Soft Matter},
edited by G. Gompper and M. Schick
(Wiley-VCH, Berlin, 2007), Vol. 3, p. 159.

\bibitem{wu:06} J. Wu,
AIChE Journal {\bf 52}, 1169 (2006).

\bibitem{harn:08} L. Harnau
Mol. Phys. {\bf 106}, 1975 (2008).

\bibitem{pars:79} J. D. Parsons,
Phys. Rev. A {\bf 19}, 1225 (1979).

\bibitem{lee:87} S. D. Lee,
J. Chem. Phys. {\bf 87}, 4972 (1987).

\bibitem{clea:96} D. J. Cleaver, C. M. Care, M. P. Allen, and M. P. Neal,
Phys. Rev. E {\bf 54}, 559 (1996).

\bibitem{samb:94} A. Samborski, G. T. Evans, C. P. Mason, and M. P. Allen,
Molec. Phys. {\bf 81}, 263 (1994).

\bibitem{camp:96} P. J. Camp, C. P. Mason, M. P. Allen, A. A. Khare, and D. A. Kofke,
J. Phys. Chem. {\bf 105}, 2837 (1996).

\bibitem{grot:96} S. C. McGrother, D. C. Williamson, and G. Jackson,
J. Phys. Chem. {\bf 104}, 6755 (1996).

\bibitem{melg:08} M. Franco-Melgar, A. J. Haslam, and G. Jackson,
Molec. Phys. {\bf 106}, 649 (2008).

\bibitem{gork:07a} M. V. Gorkunov, F. Giesselmann, J. P. F. Lagerwall,
T. J. Sluckin, and M. A. Osipov,
Phys. Rev. E {\bf 75}, 060701(R) (2007).

\bibitem{gork:07b} M. V. Gorkunov, M. A. Osipov, J. P. F. Lagerwall, and 
F. Giesselmann,
Phys. Rev. E {\bf 76}, 051706 (2007).

\bibitem{harn:02} L. Harnau and S. Dietrich,
Phys. Rev. E {\bf 65}, 021505 (2002).

\bibitem{roij:95} R. van Roij, P. Bolhuis, B. Mulder, and D. Frenkel,
Phys. Rev. E {\bf 52}, 1277 (1995).

\bibitem{graf:99} H. Graf and H. L\"{o}wen,
Phys. Rev. E {\bf 59}, 1932 (1999).

\bibitem{kram:00} E. M. Kramer and J. Herzfeld,
Phys. Rev. E {\bf 61}, 6872 (2000).

\bibitem{egge:09} E. Eggen, M. Dijkstra, and R. van Roij,
Phys. Rev. E {\bf 79}, 041401 (2009).

\bibitem{mcla:64} E. McLaughlin, M. A. Shakespeare, and A. R. Ubbelohde,
Trans. Faraday Soc. {\bf 60}, 25 (1964).

\bibitem{dogi:06} Z. Dogic and S. Fraden,
in \textit{Soft Matter},
edited by G. Gompper and M. Schick
(Wiley-VCH, Berlin, 2006), Vol. 2, p. 69.

\bibitem{kooi:01a} F. M. van der Kooij, D. van der Beek, and H. N. W. Lekkerkerker,
J. Phys. Chem. B {\bf 105}, 1696 (2001).

\bibitem{bier:04} M. Bier, L. Harnau and S. Dietrich,
Phys. Rev. E {\bf 69}, 021506 (2004).

\bibitem{hess:93} V. Hessel, H. Ringsdorf, R. Festag, and J. H. Wendorff,
Makromol. Chem. Rapid Commun. {\bf 14}, 707 (1993).

\bibitem{kokk:90} A. Kokkinia and C. M. Paleos, 
Mol. Cryst. Liq. Cryst. {\bf 186}, 239 (1990).

\bibitem{full:96} S. Fuller, N. N. Shinde, and G. J. T. Tiddy,
Langmuir {\bf 12}, 1117 (1996).

\bibitem{bier:05} M. Bier, L. Harnau and S. Dietrich,
J. Chem. Phys. {\bf 123}, 114906 (2005).

\bibitem{bier:06} M. Bier, L. Harnau and S. Dietrich,
J. Chem. Phys. {\bf 125}, 184704 (2006).

\bibitem{harn:00} L. Harnau and P. Reineker, 
J. Chem. Phys. {\bf 112}, 437 (2000). 

\bibitem{harn:01} L. Harnau, D. Costa, and J.-P. Hansen,
Europhys. Lett. {\bf 53}, 729 (2001).

\bibitem{harn:02a} L. Harnau and J.-P. Hansen, 
J. Chem. Phys. {\bf 116}, 9051 (2002). 

\bibitem{rebe:05} L. P. N. Reblo, J. N. Canongia Lopes, 
J. M. S. S. Esperanca, and E. Filipe,
J. Phys. Chem. B {\bf 109}, 6040 (2005).

\bibitem{beta:09} M. Martin-Betancourt, J. M. Romero-Enrique, and 
L. F. Rull,
J. Phys. Chem. B {\bf 111}, 9046 (2009).
\end{thebibliography}
\end{document}